\documentclass[sigconf]{acmart}

\copyrightyear{2025}
\acmYear{2025}
\setcopyright{acmlicensed}
\acmConference[UbiComp Companion '25] {Companion of the the 2025 ACM International Joint Conference on Pervasive and Ubiquitous Computing}{October 12--16, 2025}{Espoo, Finland.}
\acmBooktitle{Companion of the the 2025 ACM International Joint Conference on Pervasive and Ubiquitous Computing (UbiComp Companion '25), October 12--16, 2025, Espoo, Finland}
\acmISBN{979-8-4007-1477-1/25/10}
\acmDOI{10.1145/3714394.3754356}

\usepackage{graphicx}
\usepackage{subfigure}

\begin{document}

\title{Harmony: A Human-Aware, Responsive, Modular Assistant with a Locally Deployed Large Language Model}

\author{Ziqi YIN}
\email{yinziqi2001@toki.waseda.jp}
\affiliation{%
  \institution{Waseda University}
  \state{Tokyo}
  \country{Japan}
}

\author{Mingxin Zhang}
\email{m.zhang@hapis.k.u-tokyo.ac.jp}
\affiliation{%
  \institution{The University of Tokyo}
  \state{Tokyo}
  \country{Japan}}

\author{Daisuke Kawahara}
\email{dkw@waseda.jp}
\affiliation{%
  \institution{Waseda University}
  \state{Tokyo}
  \country{Japan}}

\renewcommand{\shortauthors}{Yin et al.}

\begin{abstract}
 Large Language Models (LLMs) offer powerful capabilities for natural language understanding, enabling more intelligent smart home assistants. However, existing systems often rely on cloud-based LLMs, raising concerns around user privacy and system dependency on external connectivity. In this work, we present \textbf{Harmony}, a privacy-preserving and robust smart home assistant powered by the locally deployable Llama3-8B model. Beyond protecting user data, Harmony also addresses reliability challenges of smaller models, such as hallucination and instruction misinterpretation, through structured prompting and modular agent design. Experimental results in both virtual environments and user studies show that Harmony achieves performance comparable to GPT-4-based systems, while enabling offline, proactive, and personalized smart home interaction.

\end{abstract}

\ccsdesc[500]{Human-centered computing~Ubiquitous and mobile computing systems and tools}
\ccsdesc[300]{Human-centered computing~Ubiquitous and mobile computing design and evaluation methods}
\ccsdesc[300]{Computing methodologies~Natural language processing}

\keywords{Smart home, Large language models, Edge computing, Privacy-preserving systems, Natural language interaction}


\maketitle

\section{Introduction}
The integration of Large Language Models (LLMs) into smart home assistants has enabled systems to interpret natural and often ambiguous commands—for example, “make the living room comfortable”—and take appropriate actions. Most of these systems leverage high-performance models like GPT-4~\cite{gpt4technicalreport} via cloud APIs, where user inputs and contextual data (e.g., sensor states, schedules, and preferences) are sent to external servers for processing.

However, this paradigm raises several critical concerns. First, the transmission of sensitive, fine-grained data to third-party servers introduces serious privacy risks. Second, the reliance on remote servers reduces robustness—loss of connectivity or API limits can severely degrade system availability. Third, commercial LLMs are often closed-source, expensive, and legally restricted, limiting transparency and reproducibility. These issues underscore the need for a more secure, reliable, and user-controllable alternative.


To address these challenges, we present \textbf{Harmony}, a smart home assistant powered by a lightweight, open-weight LLM (Llama3-8B) running entirely on local hardware. Harmony aims to maintain the natural language understanding and contextual reasoning capabilities of state-of-the-art LLMs, while enhancing data privacy, operational robustness, and user adaptability. It is designed to proactively respond to environmental cues and user patterns, even in the absence of explicit commands.

Technically, Harmony adopts a modular architecture that separates user intent recognition, decision-making, and execution. By combining structured prompting (e.g., ReAct~\cite{react}, Chain-of-Thought~\cite{cot}) with rule-based post-processing, Harmony overcomes key limitations of small LLMs such as hallucination, instruction forgetting, and inconsistent formatting~\cite{smallerllmmorehallucination}. This makes it practical for on-device deployment in real-world smart home scenarios.

This paper aims to explore two key research questions (RQ):
\begin{itemize}
    \item RQ1. What are the obstacles to applying small models to smart homes?
    \item RQ2. Can smart home assistants proactively understand and respond to users' implicit needs without speaking rather than relying solely on user instructions?
\end{itemize}
\section{Related Work}
\subsection{LLMs for Smart Home Assistants}

Recent systems such as Sasha~\cite{sasha} demonstrate how GPT-based agents can interpret complex commands and automate smart home actions with greater flexibility than traditional rule-based platforms~\cite{ifttt}  or voice assistants. However, most LLM-powered assistants rely on cloud-based models like GPT-4~\cite{gpt4technicalreport}, requiring sensitive user data to be transmitted externally, raising well-known privacy risks~\cite{EkmanubiquitousPrivacy, ShiEdgecomputingPrivacy}. Additionally, such reliance introduces latency and limits customizability. To address this, we explore the use of locally deployable lightweight models (e.g., Llama3-8B). Yet these models suffer from hallucination and instruction forgetting~\cite{surveyhallucination, smallerllmmorehallucination, miaoAttentionClosely, chenAttentionClosely}, making robust prompting and architecture design essential.

\subsection{Prompting Techniques and LLM-based Agents}

Prompt engineering techniques have been widely explored to improve the reliability and interpretability of LLM behavior. Methods such as Chain-of-Thought (CoT)~\cite{cot} and ReAct~\cite{react} explicitly structure reasoning steps to guide decision-making, which has proven helpful in reducing hallucination and improving alignment with user intent.

LLM-based agents are designed to perceive their environment, reason through context, and take autonomous actions~\cite{surveyllm, surveyagent}. In the context of smart homes, such agents are expected not only to understand commands but to proactively infer user needs based on historical interaction data and sensor inputs. This requires a delicate balance between flexibility, accuracy, and interpretability—especially when using small-scale models in privacy-sensitive environments.

\begin{figure*}[t]
    \centering
    \includegraphics[width=0.8\linewidth]{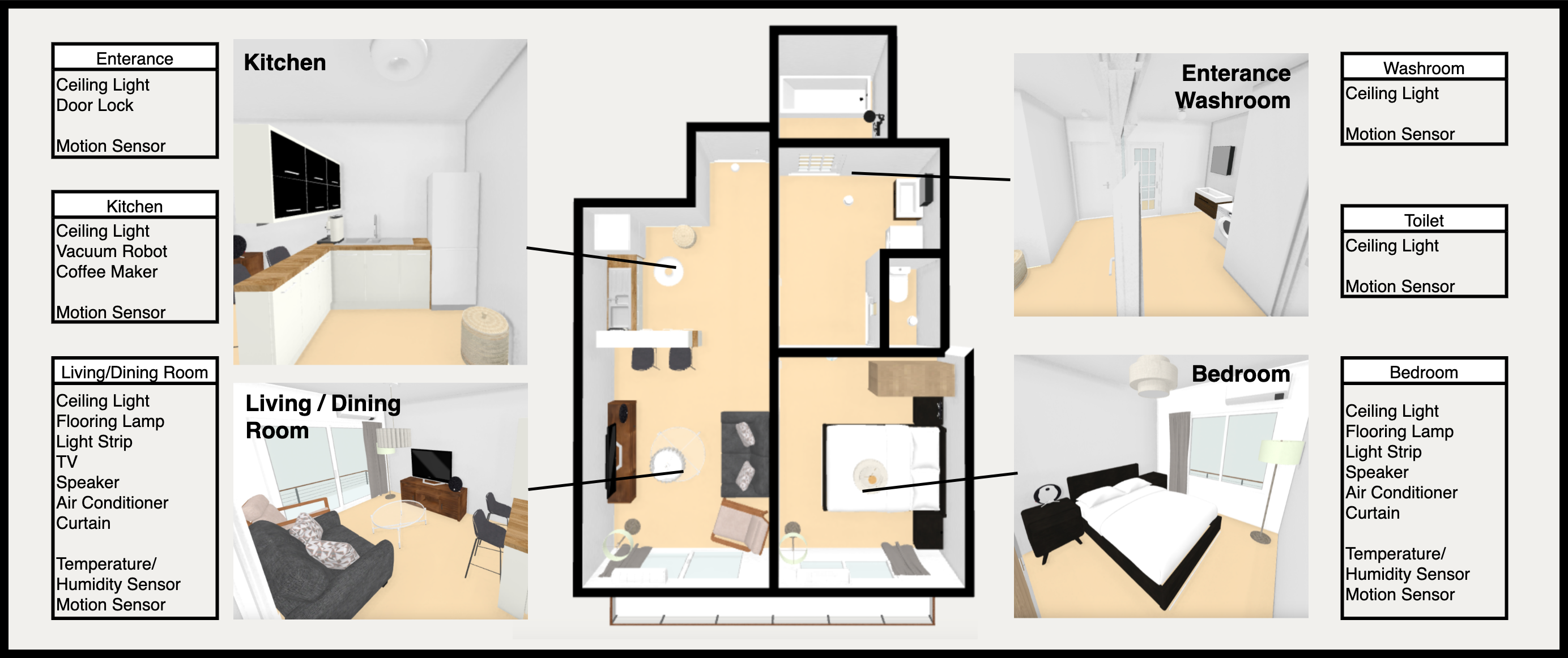}
    \caption{Home Template}
    \label{fig:hometemplate}
\end{figure*}

\section{Method}
\subsection{Challenges of Lightweight LLMs in Smart Home Scenarios}
Prior research indicates that smaller LLMs are more prone to hallucination and formatting errors. To quantify this in a smart home context, we replicated Sasha’s framework using CoT prompts and tested both GPT-4 and Llama3-8B on a set of IFTTT routines in a richly equipped virtual environment (Figure.~\ref{fig:hometemplate}).

\begin{table}[h]
    \centering
    
    \caption{Preliminary Experimental Result}
    \begin{tabular}{c|ccc}
    \hline
         Model & Device FP & Room FP & FN \\
        \hline
          Llama3-8B   & 7&1 & 6\\
          GPT-4   & 0&0 & 4 \\
        \hline
    \end{tabular}
    \label{tab:hallunicationExperiment}
\end{table}



As shown in Table~\ref{tab:hallunicationExperiment}, Llama3-8B exhibits significantly higher error rates, including hallucinated devices and excessive strictness. Inconsistent JSON formatting is also observed in Llama3-8B.
Given these findings, we have identified several core challenges in using smaller LLMs in smart space scenarios to answer RQ1:
\begin{itemize}
    \item More severe hallucinations.
    \item Difficulty in following instructions.
    \item Inability to generate correctly formatted JSON without specialized fine-tuning.
\end{itemize}
These limitations highlight the need for techniques that can enhance the robustness of small LLMs in context-rich, real-time environments. The goal of our work is to design prompting strategies and system-level safeguards that enable such models to perform reliably and safely, which form the core of our proposed Harmony framework.



\subsection{Harmony Architecture}
To improve the reliability of small models like Llama3-8B in the smart home setting, we designed prompting strategies tailored to address hallucination, instruction-following issues, and formatting instability. Our two key techniques are:

\textbf{1. Structured, localized reasoning.}
Unlike Sasha’s simple three-stage chain of thought, we strictly limited the LLM's response range without compromising its reasoning ability. We adjusted our prompts to the model based on its responses. Thus, the model always finds clear instructions at the end of the prompt, leading to more compliant actions.

\textbf{2. ReAct-style context grounding.}
To further enhance decision quality, we incorporated ReAct-style prompting, where the model must first explain its interpretation of the user’s situation before generating actions. This encourages deeper grounding and reduces overconfidence in vague scenarios—an issue commonly observed in smaller models.

\begin{figure*}[t]
  \includegraphics[width=1.9\columnwidth]{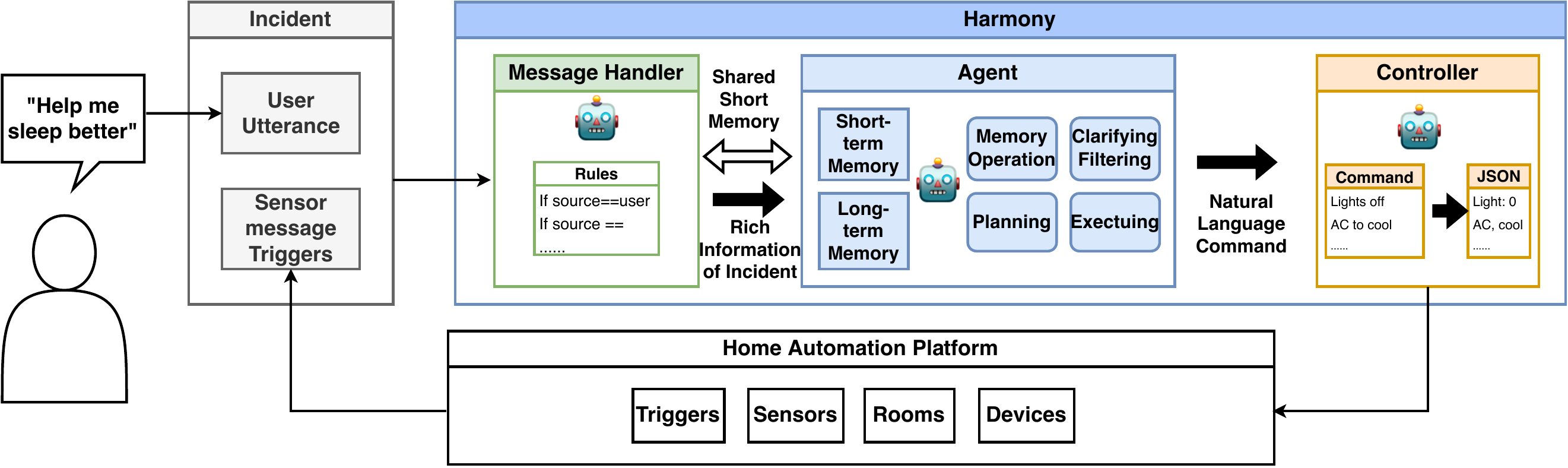}
  \caption{Illustration of Harmony}
  \label{fig:harmony_arch}
\end{figure*}

Based on the above content, we propose the Harmony framework, shown in Figure~\ref{fig:harmony_arch}. Unlike smart home assistants such as Sasha, Harmony adopts a design that relies solely on small LLMs that can run locally. This framework can be divided into three parts, each designed with detailed steps to effectively suppress hallucinations in small-scale LLMs and promote correct behavior and proactive user service. These three parts are the Message Handler, Agent, and Controller.

\textbf{Message Handler}
The Message Handler assesses the urgency of events and infers user intent using short-term memory shared with the Agent. It interprets sensor data in context to a real world activity. This inference will be passed to the Agent for further action.

\textbf{Agent}
The Agent manages appliances based on inferred user needs. It updates memory or takes action depending on the message. If the message requires action, it checks feasibility and either runs a scenario or controls devices directly. Persistent actions are stored as rules. After execution, it summarizes the interaction for future reference.

\textbf{Controller}
The Controller converts the Agent’s task plans into JSON commands for the smart home system. It uses rule-based logic and LLMs, parsing structured plans and assigning actions to specific rooms.

\section{Virtual Environment Experiment}
In the virtual environment experiment, we used a highly diverse smart home environment containing lights, security, environment controlling and entertainment devices, a vacuum robot, and extra devices, as shown in Figure~\ref{fig:hometemplate}. Each room is equipped with at least one light and motion sensor. Climate sensors are also equipped in living areas. We used IFTTT routines for virtual environment evaluation by recording the assistant decision of each instruction in the IFTTT routines for automatic and manual evaluation.

\subsection{Automatic Evaluation}
\subsubsection{Method}
We used Sasha's relevance evaluation method, which focuses solely on assessing the executability of the task plans generated by the smart home assistant without considering subjective impressions of task plans. 
We used IFTTT routines for evaluating False Positive, False Negative, and Accuracy.

    
    
    

\begin{table}
    \centering
        \caption{Automatic Evaluation Result}
    \begin{tabular}{c|ccc}
    \hline
          & False Positive & False Negative  & Accuracy\\
        \hline
         Harmony   & 0.08&0 & 0.92\\
          Sasha (GPT-4)   & 0.10&0 & 0.90\\
          Sasha (Llama3-8b) & 0.19 & 0.18 & 0.63\\
        \hline
    \end{tabular}

    \label{virtualExperiment}
\end{table}
\subsubsection{Result}
The result is shown in Table~\ref{virtualExperiment}. Harmony and Sasha (GPT-4) perform equally well in false positive and accuracy, while Sasha (Llama3-8b) has a higher false positive and false negative rate, resulting in a lower accuracy.
\subsection{Manual Evaluation}
\subsubsection{Method}
We employed a questionnaire-based survey method. Each questionnaire consisted of multiple items, and each item was composed of two parts: the first part presented a user instruction representative of an IFTTT routine (e.g., ``Turn on the light for me''), and the second part showed the corresponding assistant's action. As the study involved three different assistants, three separate questionnaires were randomly presented to participants without offering the assistant's name. 

In the questionnaire, participants were asked to take the user's perspective. They then rated the assistant's action on a 5-point Likert scale, evaluating how well the action fulfilled the instruction. We collected questionnaire data from 18 participants and conducted a repeated measure ANOVA to compare the performance across the three assistants.

\begin{figure}
    \centering
    \includegraphics[width=0.8\columnwidth, trim={0 25 0 0}, clip]{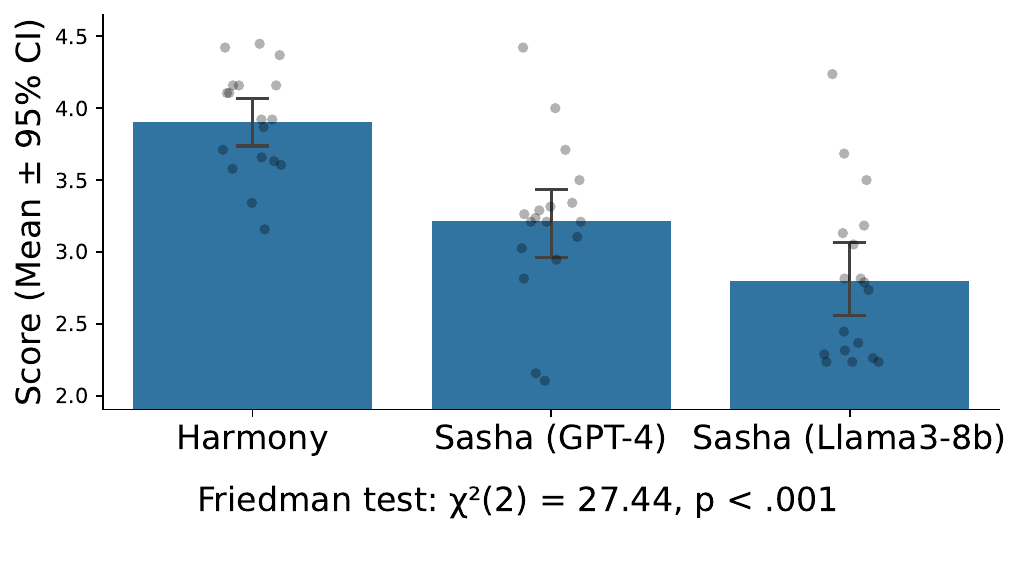}
    \caption{Result of questionnaire-based survey}
    \label{fig:questionnaire}
\end{figure}
\subsubsection{Result}
The evaluation results of Harmony and the two baseline assistants are presented in Figure \ref{fig:questionnaire}.
Statistical analysis revealed a significant effect of assistant on user satisfaction ($\chi^2$(2)=27.44, p < .001).
Among the three, Harmony received the highest average score (3.91±0.37), significantly outperforming both Sasha (GPT‑4) (3.21±0.55, p < .001) and Sasha (Llama3‑8b) (2.80±0.58, p < .001).
These findings suggest that Harmony can deliver a user experience that not only rivals but even exceeds that of GPT‑4-based assistants, thanks to effective prompting and task adaptation.

\section{User Study}
\subsection{User Study Procedure}
We constructed the experimental environment with reference to Figure~\ref{fig:hometemplate}. A total of N = 5 participants were recruited via word-of-mouth and social media to assess the interaction capabilities of Harmony across several daily life scenarios. Each experimental session lasted approximately 90–120 minutes.
\begin{table}
    \centering
        \caption{Demographic characteristics of user study experiment participants}
    \begin{tabular}{c|ccc}
    \hline
         ID &  Age & Education& Smart Home Familiarity\\
        \hline
         P1   & 18-25& Graduate& Very\\
          P2   & 26-35& Bachelor & Moderately\\
          P3 & 26-35 & Doctor &Moderately\\
          P4   & 18-25& Bachelor& Not at all\\
          P5   & 18-25& Graduate& Slightly\\
        \hline
    \end{tabular}

    \label{rwplist}
\end{table}
Each participant completed the experiment following these steps:
\begin{enumerate}
  \item \textbf{Introduction and Demonstration}: A briefing was provided on the procedures and usage instructions.
  \item \textbf{Execution of Three Preset Scenarios}: Returning Home, Morning Routine, and Hosting a Party.
  \item \textbf{Selection and Execution of a Customized Scenario}.
  \item \textbf{Interview Session}: Participants shared their experiences and reflections on the interaction process.
\end{enumerate}
Participants were asked to design a scenario based on their own daily routines and interact with Harmony until they felt the assistant had adequately satisfied their goal. Throughout all scenarios, participants were allowed to interact with Harmony using any kind of command and were encouraged to provide comment in real-time.

\subsection{Results}
\begin{figure}[htbp]
  \centering
  \subfigure{\includegraphics[width=1\columnwidth]{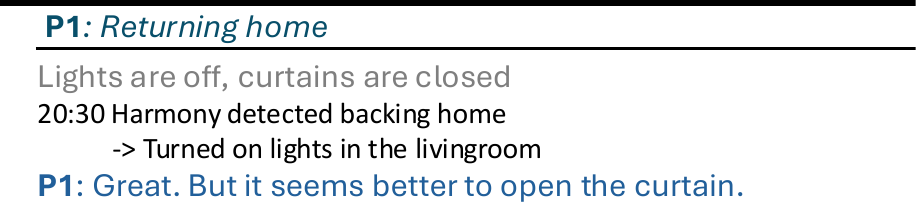}}
  \vspace{-5pt}
\subfigure{\includegraphics[width=1\columnwidth]{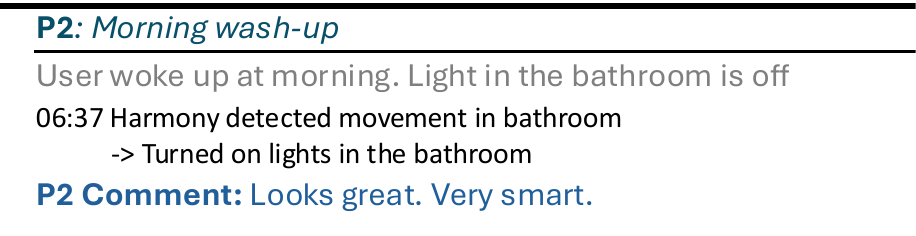}}
\subfigure{\includegraphics[width=1\columnwidth]{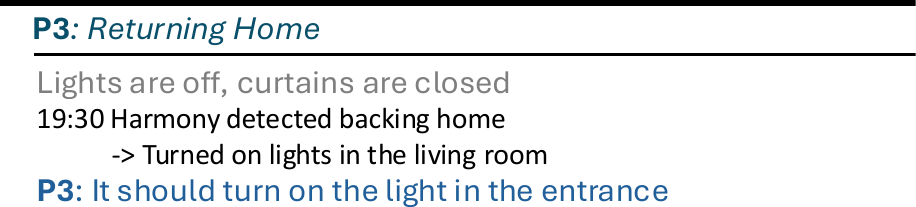}}
\subfigure{\includegraphics[width=1\columnwidth]{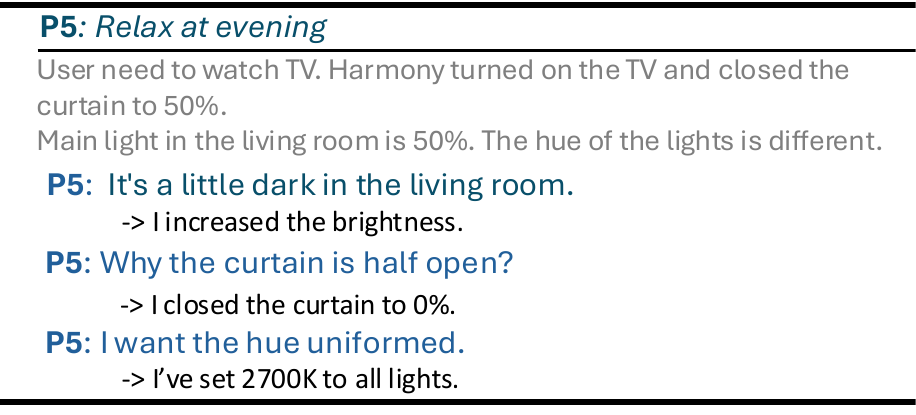}}
\caption{Examples of user study}
  \label{fig:userstudy}
\end{figure}

We present common themes from user study in Figure~\ref{fig:userstudy}. Headers indicate the participant, followed by the scenario that the interaction occurred in.

We observed three points to answer RQ2:
\begin{enumerate}
    \item \textbf{Proactive Assistance Enhances User Satisfaction:}
When the assistant offered help proactively through commonsense reasoning on sensor status change, user experience significantly improved. particularly among participants who were initially skeptical about voice-based interaction.
    \item \textbf{Understanding User Preferences is Crucial:}
Failure to accurately infer user preferences led to noticeable dissatisfaction, especially when users issued ambiguous commands.
    \item \textbf{Interaction Logs as Preference Knowledge:}
Interaction histories accumulated over multiple sessions helped the system better adapt to individual user preferences.
\end{enumerate}

\section{Limitation and Future Work}
While Harmony achieves promising results, several limitations remain. The inference capabilities of the local LLM (Llama3-8b) are still unstable in complex scenarios, and the system lacks a robust mechanism for learning and adapting to user preferences over time. Future work will explore continuous learning approaches to personalizing behavior and improve contextual understanding. We also plan to expand the scope of experiments with more diverse users and environments to validate scalability and generalizability.

\section{Conclusion}
This paper presented Harmony, a smart home assistant powered by a locally deployed lightweight LLM. By combining structured prompting and a modular agent architecture, Harmony achieves privacy-preserving, proactive assistance without relying on cloud computation. Both virtual and user evaluations suggest that Harmony can deliver a satisfying and responsive experience, comparable to GPT-4-based systems. Our results highlight the potential of local LLMs in everyday environments, opening new directions for secure and personalized ubiquitous computing.
\begin{acks}
This work was partially supported by JSPS KAKENHI Grant Number JP24H00727.
\end{acks}

\bibliographystyle{ACM-Reference-Format}



\end{document}